\documentclass[aps,prl,article,twocolumn,showpacs,preprintnumbers,amsmath,amssymb,superscriptaddress]{revtex4}
\date{\today}
\usepackage{epsfig}
\usepackage{subfigure}
\usepackage{graphicx}
\usepackage{dcolumn}
\usepackage{bm}
\usepackage[colorlinks,linkcolor=blue,hyperindex,CJKbookmarks]{hyperref}
\usepackage{float}
\usepackage{hyperref}
\usepackage{comment}
\begin{document}

\title{Light-Enhanced Spin Fluctuations and $d$-Wave Superconductivity at a Phase Boundary}
\author{Yao Wang }
 \affiliation{Department of Applied Physics, Stanford University, Stanford, California 94305, USA}
 \affiliation{SLAC National Accelerator Laboratory, Stanford Institute for Materials and Energy Sciences, 2575 Sand Hill Road,
Menlo Park, California 94025, USA}
\affiliation{Department of Physics, Harvard University, Cambridge, Massachusetts 02138, USA}
\author{Cheng-Chien Chen}%
\affiliation{Department of Physics, University of Alabama at Birmingham, Birmingham, Alabama 35294, USA}
\author{B. Moritz}
\affiliation{SLAC National Accelerator Laboratory, Stanford Institute for Materials and Energy Sciences, 2575 Sand Hill Road,
Menlo Park, California 94025, USA}%
\affiliation{Department of Physics and Astrophysics, University of North Dakota, Grand Forks, North Dakota 58202, USA}
\author{T. P. Devereaux}
 \email[Author to whom correspondence should be addressed to Y.W. (\href{mailto:yaowang@g.harvard.edu}{yaowang@g.harvard.edu}) or T.P.D. (\href{mailto:tpd@stanford.edu}{tpd@stanford.edu})
]{}
\affiliation{SLAC National Accelerator Laboratory, Stanford Institute for Materials and Energy Sciences, 2575 Sand Hill Road,
Menlo Park, California 94025, USA}%
\affiliation{Geballe Laboratory for Advanced Materials, Stanford University, Stanford, California 94305, USA}
\date{\today}
\begin{abstract}
Time-domain techniques have shown the potential of photo-manipulating existing orders and inducing new states of matter in strongly correlated materials. Using time-resolved exact diagonalization, we perform numerical studies of pump dynamics in a Mott-Peierls system with competing charge and spin density waves. A light-enhanced $d$-wave superconductivity is observed when the system resides near a quantum phase boundary. By examining the evolution of spin, charge and superconducting susceptibilities, we show that a sub-dominant state in equilibrium can be stabilized by photomanipulating charge order to allow superconductivity to appear and dominate. This work provides an interpretation of light-induced superconductivity from the perspective of order competition, and offers a promising approach for designing novel emergent states out of equilibrium. 
\end{abstract}
\maketitle

With the development of pump-probe instruments in recent years, time-domain techniques have been widely applied to the study of complex quantum materials\cite{zhang2014dynamics, giannetti2016ultrafast}. The rich information revealed in the extra time dimension holds potential to characterize ordered states\cite{hellmann2012time, kim2012ultrafast, patz2014ultrafast, dal2014snapshots, rubbo2011resonantly, carrasquilla2013scaling}, disentangle different variables\cite{lee2012phase, schachenmayer2013entanglement, wang2016using}, and trace pathways of electronic evolution from designed perturbations\cite{PhysRevLett.92.237401, reed2010effective, wang2014real}. On the one hand, the electronic or structural properties can be transiently engineered by the pump field, which could potentially stabilize new states of matter or drive incipient phase transitions\cite{schmitt2008transient, fausti2011light, liu2012terahertz,  balzer2015nonthermal, mitrano2016possible}. On the other hand, the development of these phenomena and their subsequent relaxation further reveal information about the underlying physics\cite{PhysRevLett.102.106405, tomeljak2009dynamics, lu2012enhanced, rincon2014photoexcitation, lu2015photoinduced}. Along with the achievements of ultrafast experiments, the development of nonequilibrium photomanipulation in microscopic theories is in demand.

In particular, nonequilibrium studies of strongly correlated materials are complicated due to intertwined orders. In such systems, the microscopic mechanisms of intriguing emergent phenomena such as unconventional superconductivity remain elusive\cite{kivelson2003detect, fujita2011progress, davis2013concepts}. Therefore, ultrafast techniques could be helpful because of their capability to shift the balance between different emergent phases and create new states of matter inaccessible in equilibrium\cite{rini2007control, gedik2007nonequilibrium, gorshkov2011tunable, li2013femtosecond, hazzard2013far}. Efforts have been made to enhance superconductivity\cite{fausti2011light, kaiser2014optically, nicoletti2014optically, casandruc2015wavelength, mitrano2016possible}, while attempts towards an understanding have been made via phenomenological and mean-field theories\cite{patel2016light, sentef2016theory}. A microscopic understanding, however, remains open due to the lack of adequate treatment of the strongly coupled degrees of freedom in correlated electron systems. It is significant for the prediction and design of superconducting states in complex materials to unravel whether the enhanced superconductivity is a new state born from an underlying instability or a result of a photomanipulation of balanced phases.

Previous theoretical studies in both equilibrium and time-domain have shown the intimate relationship between spin, charge and lattice variables with superconductivity\cite{kivelson2003detect, abbamonte2005spatially, fujita2011progress, ghiringhelli2012long, keimer2015quantum, davis2013concepts, peronaci2015transient, murakami2016multiple, sentef2016theory}. For example, $s$-wave superconductivity can be induced by an interaction quench in a strongly correlated system\cite{bittner2017light}. Therefore, the emergence of $d$-wave superconductivity in correlated electrons may naturally lie in the manipulation of different intertwined variables. For this purpose, we perform numerical studies of a pumped Mott-Peierls system with gapped charge-density-wave (CDW) and spin-density-wave (SDW) orders using time-resolved exact diagonalization. During the pump, we find photoenhanced superconductivity in the vicinity of the phase boundary from the Peierls state, while there is no apparent change of the superconducting order parameter deep in either the gapped Peierls or Mott phases. The photoenhanced remnant pairing instabilities increase and become divergent near the phase boundary. Through the comparison with $d$-wave projected spin fluctuations, we attribute such a substantial enhancement to photoinduced spin excitations, in contrast to quasiparticle weight and bandwidth engineering. This work thereby provides a novel perspective on creating nonequilibrium emergent phenomena, particularly superconductivity, through the control of the interactions near a quantum phase transition.

In order to simulate the competition between Peierls/CDW and Mott/SDW phases, we adopted a two-dimensional Peierls-Hubbard model, which describes the lattice degrees of freedom by a uniform dimerization. The model Hamiltonian reads $\mathcal{H}= \mathcal{H}_{\textit{e-e}}+ \mathcal{H}_{\textit{e-ph}}$\cite{Hubbard,Holstein, wang2016using}:
\begin{eqnarray}
\mathcal{H}_{\textit{e-e}} &=&-t_h\sum_{\langle\textbf{i},\textbf{j}\rangle,\sigma}(c_{\textbf{i}\sigma}^\dagger c_{\textbf{j}\sigma}+h.c.)+U\sum_\textbf{i} n_{\textbf{i}\uparrow}n_{\textbf{i}\downarrow}\nonumber\\
\mathcal{H}_{\textit{e-ph}}&=&-\frac{g}{\sqrt{N}}(b^\dagger +b)\sum_{\textbf{i},\sigma}(-1)^{i_x+i_y}n_{\textbf{i}\sigma}+\Omega\, b^\dagger b
\end{eqnarray}
where $t_h$ is the nearest-neighbor hopping integral, $c_{\textbf{i}\sigma}^\dagger$ ($c_{\textbf{i}\sigma}$) and $n_{\textbf{i}\sigma}$ are the electron creation (annihilation) and number operators at site $\textbf{i}$ of spin $\sigma$, $U$ is the on-site Coulomb repulsion, and $b^\dagger$ ($b$) and $\Omega$ are the phonon creation (annihilation) operator and frequency, respectively. The dimensionless electron-electron (\textit{e-e}) and electron-phonon (\textit{e-ph}) coupling strengths are defined as $u=U/t_h$ and $\lambda=g^2/t_h\Omega$, respectively. The phonon frequency is set to $\Omega\!=\!t_h$ as in Ref.~\cite{clay2005intermediate, wang2016using}.

\begin{figure}[!t]
\begin{center}
\includegraphics[width=\columnwidth]{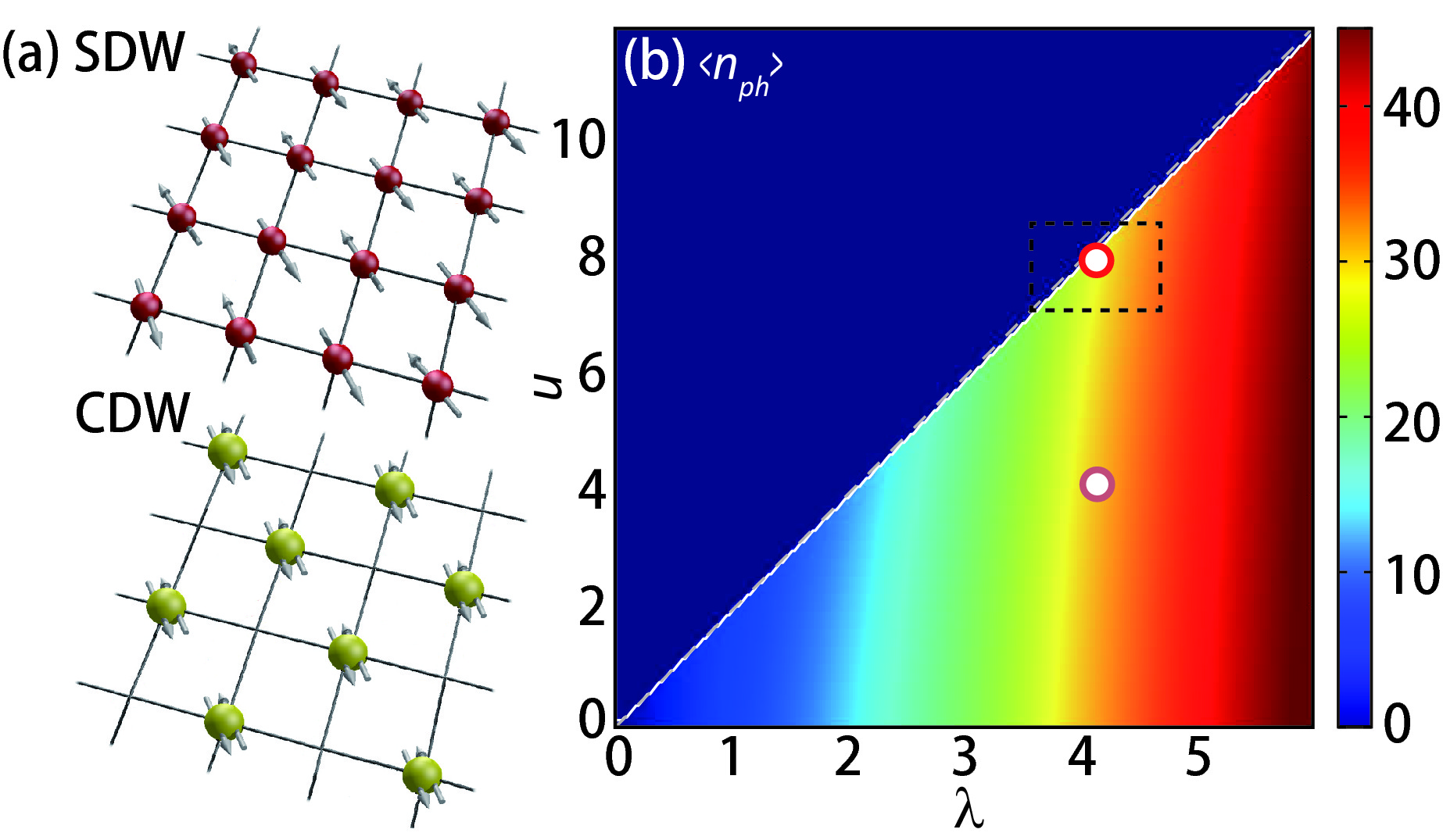}
\caption{\label{fig:1} (a) Schematics of SDW and CDW insulating states. (b) Average phonon occupancy $\langle n_{\rm ph}\rangle$ at various $\lambda$ and $u$. The dashed lines indicate the phase boundaries in the anti-adiabatic limit where $u_\textrm{eff}\!=\!0$, while the solid line tracks the numerical boundary where the translational symmetry breaks and the ground state changes from doubly-degenerate (Peierls phase) to non-degenerate. The red and orange circles denote the parameters used in Fig.~\ref{fig:2} and the boundary-crossing dotted bar shows the parameter space traversed in Fig.~\ref{fig:3}.
}
\end{center}
\end{figure}

At half-filling, this model describes the competition of CDW and SDW states at the nesting momentum $\textbf{q}=(\pi,\pi)$, which is consistent with the Hubbard-Holstein model: the presence of both \textit{e-e} and \textit{e-ph} effects leads to this competition and a metallic region between the ordered phases\cite{clay2005intermediate, fehske2008metallicity, bauer2010competing, nowadnick2012competition, murakami2013ordered, hohenadler2013excitation, greitemann2015finite}. Fig.~\ref{fig:1}(b) shows the equilibrium phase diagram as a function of $u$ and $\lambda$ in terms of phonon occupancy $\langle n_{\rm ph}\rangle$. A Peierls phase with leading checkerboard CDW order and large distortion (or phonon numbers) exists on the $u\ll 2\lambda$ side, while a Mott phase with leading SDW order lives on the other side. In contrast to the one-dimensional situation\cite{clay2005intermediate, fehske2008metallicity, hohenadler2013excitation, greitemann2015finite, wang2016using}, the intermediate metallic phase is relatively narrow at zero temperature due to the ordered SDW. The calculations are performed on square clusters of $N\!=\!8$ sites with periodic boundary conditions and maximum phonon occupation $M\!=\!127$, which is sufficient for convergence within the range of our phase diagram [see Fig.~\ref{fig:1}(b)]. We use exact diagonalization with the parallel Arnoldi method\cite{lehoucq1998arpack} to determine the equilibrium ground state wavefunction $\big|\psi(t=-\infty)\big\rangle$.

\begin{figure}[!t]
\begin{center}
\includegraphics[width=8.5cm]{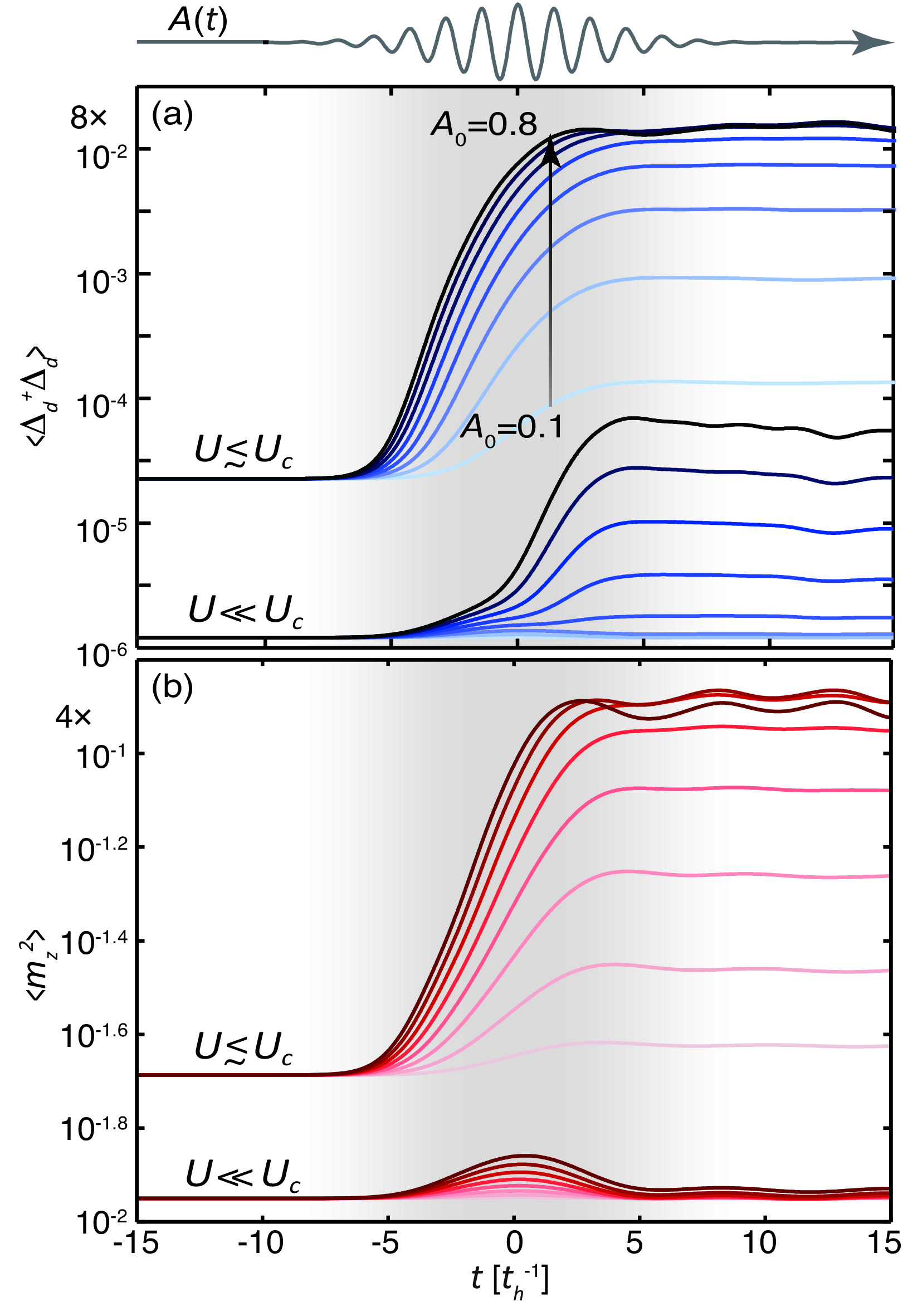}
\caption{\label{fig:2} 
Evolution of (a) $d$-wave pairing correlation $\langle \Delta_d^\dagger \Delta_d\rangle$ and (b) magnetization $\langle m_z^2\rangle$ (in log scale) during and after the pump, for systems deep [$u=3.9$, $\lambda=4$, bottom] and shallow [$u=7.8$, $\lambda=4$, top] in the Peierls phase, respectively. The darkness of curves denotes the pump strength varying from $A_0=0.1$ to 0.8.  
}
\end{center}
\end{figure}

\begin{figure*}[!ht]
\begin{center}
\includegraphics[width=16cm]{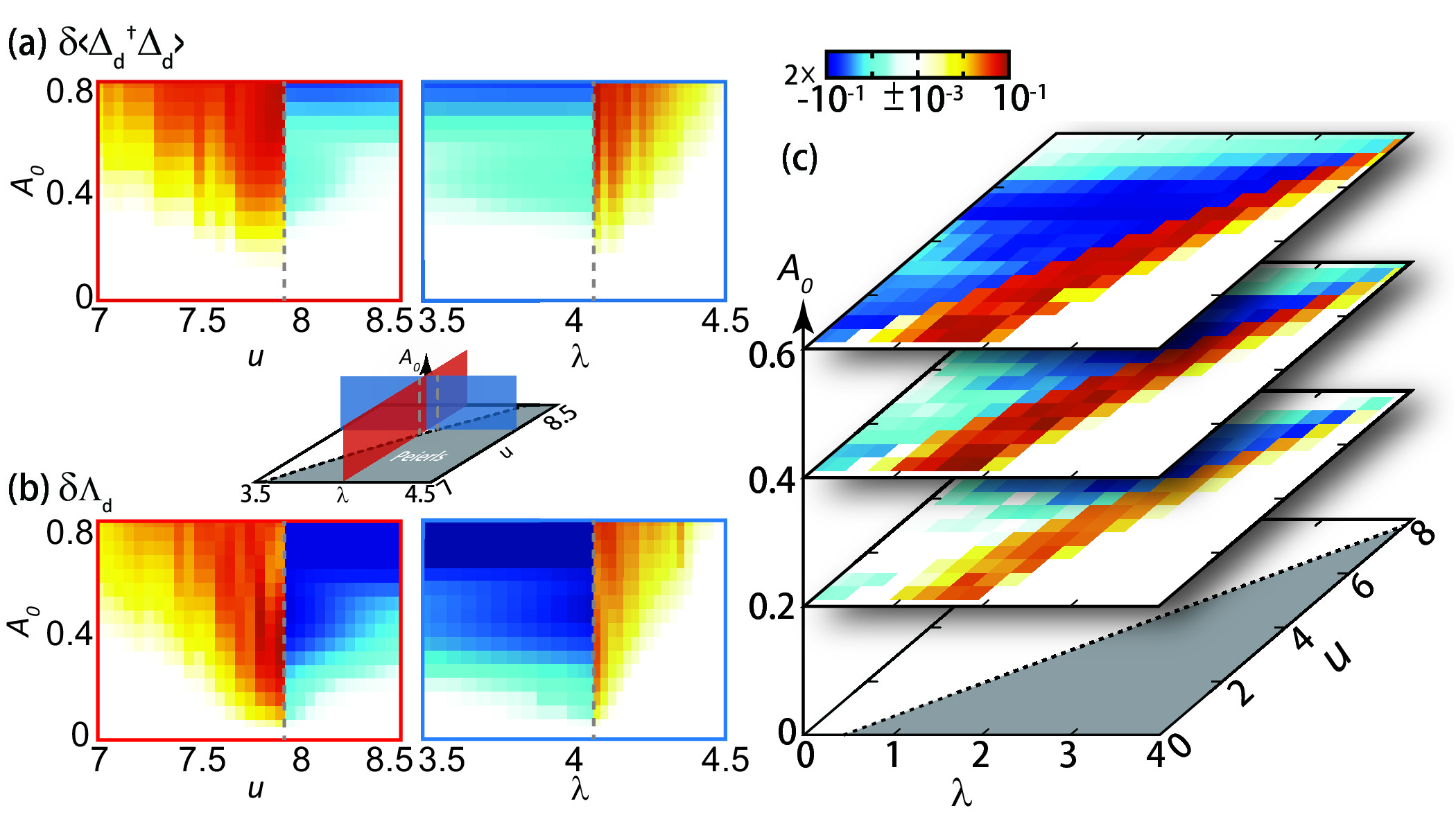}
\caption{\label{fig:3} Change of (a) pairing correlation and (b) $d$-wave projected spin fluctuations evaluated in the post-pump state ($t=10t_h^{-1}$) for various pump strength $A_0$ and parameter sets near the phase boundary along two parameters. The cut positions are denoted by the color plains in the inset. (c) The change of pairing correlation with fixed pump strengths, over a wider range of interaction parameter sets. The colormaps are plotted in logarithmic scale. The dashed lines indicate the phase boundary.
}
\end{center}
\end{figure*}

In the presence of an external field, a time-dependent Hamiltonian $\mathcal{H}(t)$ can be written with the Peierls substitution $c_{\textbf{i}\sigma}^\dagger c_{\textbf{j}\sigma} \rightarrow e^{i\mathbf{A}(t)\cdot (\bf r_j - r_i)} c_{\textbf{i}\sigma}^\dagger c_{\textbf{j}\sigma}$. Here we use an oscillatory gaussian vector potential in the temporal gauge to simulate a pulsed laser pump [see Fig.~\ref{fig:2}]
\begin{eqnarray}
\mathbf{A}(t)=A_0 e^{-(t-t_0)^2/2\sigma_t^2}\cos[\omega_0 (t-t_0)] \mathbf{e}_{\rm pol}.
\end{eqnarray}
As the nesting momentum for both phases is $(\pi,\pi)$, we select diagonal polarization $\mathbf{e}_{\rm pol} = (\mathbf{e}_{\rm x}+\mathbf{e}_{\rm y})/\sqrt{2}$. We use the Krylov subspace technique\cite{manmana2007strongly, balzer2012krylov, park1986unitary, hochbruck1997krylov, moler2003nineteen} to evaluate the time evolution of a state $|\psi(t\!+\!\delta t)\rangle\!=\!e^{-i\mathcal{H}(t)\delta t}|\psi(t)\rangle$. Throughout this work, the pump frequency is set to be $\omega_0=4.4t_h$, which is close to the Mott gap size $\sim U-4t_h$. Further discussion on the impact of the pump frequency is given in the Supplementary Materials\cite{SUPMAT}.

The Peierls-Hubbard model correctly captures the competition and quantum phase transition of SDW and CDW states. Its nonequilibrium dynamics has been shown to reveal the critical softening of bosonic excitations, reflecting the intertwined nature of the fermion-boson coupling\cite{wang2016using}. This further motivates the present work on a two-dimensional geometry, investigating $d$-wave superconductivity. We monitor the time-dependent $d$-wave pairing correlation $\langle \Delta_d^\dagger \Delta_d\rangle$ with various pump intensities\cite{intensityNote} [see Fig.~\ref{fig:2}(a)], where $\Delta_d = \sum_{\bf k} \gamma_d(\textbf{k})c_{\mathbf{k}\uparrow}c_{\mathbf{k}\downarrow}$ and $\gamma_d(\textbf{k})=\cos k_x-\cos k_y$. A phase average of the pump pulse is adopted to filter out the phase-locked fast oscillations $\sim U$, which are not relevant here. Deep in the Peierls phase ($u\ll u_c$) the pairing correlation gradually increases with pump intensity, but the dynamics are restricted within a minimal amplitude due to the dominant Peierls phase. Considering that the effective interaction remains unchanged in the BCS picture, this small enhancement can be attributed to pump-enhancement of quasiparticle weights in a previously gapped insulator.

However, near the phase boundary, where both CDW and SDW orders are well balanced ($u\lesssim u_c$), the $d$-wave pairing displays a substantial enhancement -- by three orders of magnitude. Since the equilibrium phase on the Peierls side has small superconducting correlations, this relatively strong enhancement is related to the ground state's proximity to the Mott phase transition boundary. A simulation of the magnetization $\langle m_z^2\rangle$ reveals this potential connection [see Fig.~\ref{fig:2}(b)]: unlike the dynamics deep in the Peierls phase with no surviving magnetism, the pumped $\langle m_z^2\rangle$ displays considerable enhancement near the phase boundary, which persists following the pump. The increased $d$-wave pairing instability is thus connected to the pump-induced change of the effective spin interactions.

The rise of magnetism and superconductivity can be linked naturally to the increase of the fluctuations and associated bosonic excitations near the critical point\cite{wang2016using}. To establish this connection, we fine tune parameters near the phase boundary and examine the time evolution of the post-pump $\langle \Delta_d^\dagger \Delta_d\rangle$ for $u\simeq u_c$ [see Fig.~\ref{fig:3}(a)], tracking the susceptibilities along both the $u$ and $\lambda$ directions near the phase boundary. The Peierls side displays a ``critical fan'' of pairing correlations, which are further enhanced at larger pump fluences near the critical point. However, the pairing instability is suppressed in the Mott phase. This implies that spin fluctuations play a dominant role in the development of pairing out of equilibrium.

To demonstrate the influence of spin excitations on the pairing correlations, we further examine the $d$-wave projected spin fluctuations $\Lambda_d$. Such a projected fluctuation is claimed to mediate $d$-wave superconductivity as a pairing glue, in an RPA-like scenario\cite{tsuei2000pairing,scalapino2012acommon,maier2016pairing}. The nonequilibrium $\Lambda_d(t)$ can be obtained through the measure of the spin response functions [see Ref.~\onlinecite{wang2017producing} and the Supplementary Materials \cite{SUPMAT} for calculation details]. As shown in Fig.~\ref{fig:3}(b), approaching the Peierls phase boundary or increasing the pump field will enhance the post-pump $\Lambda_d$, although it is suppressed on the Mott side of the phase boundary. The agreement between the remnant $\Lambda_d$ and the pairing correlations reflects that spin fluctuations contribute as a pairing glue in the $d$-wave channel. Therefore, the photoenhancement of $d$-wave superconductivity near the quantum phase transition is a result of competing interactions: the charge ordered state measured by the charge structure factor $N(\pi,\pi)$ [not shown here] is reduced by the pump field, releasing spin fluctuations to pair electrons with a $d$-wave symmetry.

As the charge order proves more vulnerable to an external pump in the weak coupling limit\cite{wang2016using}, other instabilities such as spin and superconductivity are expected to emerge. Fig.~\ref{fig:3}(c) shows an extended phase diagram down to the weak coupling region, with fixed pump strengths $A_0$. From the horizontal perspective, such a ``critical fan'' increasingly opens up with the decrease of interaction parameters $u$ and $\lambda$. The region where pairing correlations could be enhanced can be 2-3 times wider than the strong-coupling regime with $u\sim 8$. However, the maximum enhancement that can be achieved by a given pump remains roughly the same. The fact that $\langle \Delta_d^\dagger \Delta_d\rangle$ contour follows the phase boundary, rather than the contour of CDW orders in Fig.~\ref{fig:1}(b), reflects that it is influenced more by the buildup of spin fluctuations than by simply melting the CDW order. In contrast to the tiny enhancement deep in the phase, the photomanipulation of effective interactions instead of quasiparticle weight dominates near the critical point.

\begin{figure}[!t]
\begin{center}
\includegraphics[width=9cm]{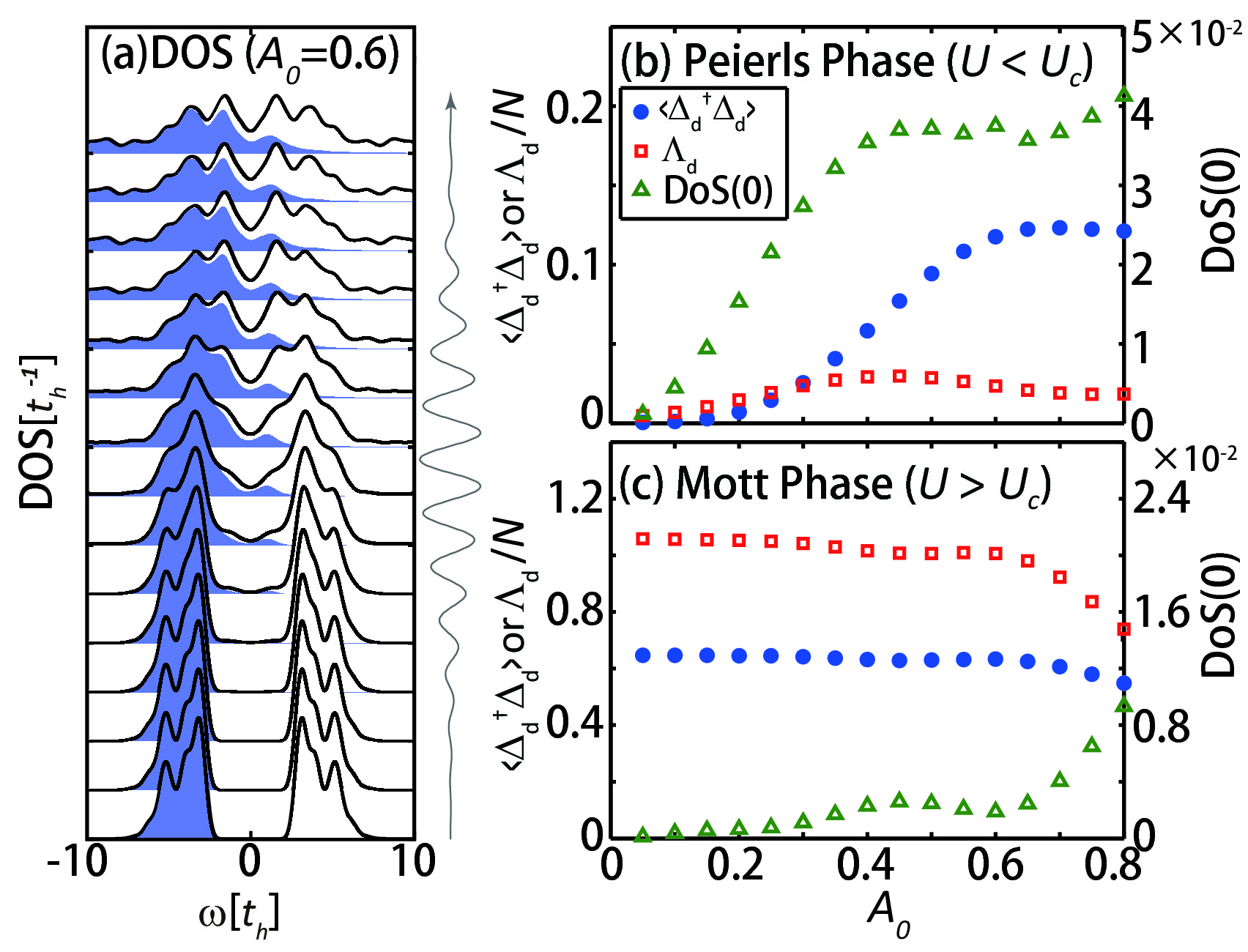}
\caption{\label{fig:4} (a) Evolution of DOS during the pump. The shaded regions denote occupied states. (b,c) Comparison of post-pump ($t=10\,t_h^{-1}$) pairing correlation (solid blue circles), spin fluctuations (red squares) and quasiparticle weight (green triangles) as a function of pump strength from (b) Peierls phase ($u=7.8$) and (c) Mott phase ($u=8$). The quasiparticle weights are plotted against the right axis.
}
\end{center}
\end{figure}

In contrast to the Peierls phase where the maximal enhancement appears near the middle or end of the pump, the nonequilibrium pairing correlation $\langle \Delta_d^\dagger \Delta_d\rangle$ is enhanced only at the very beginning of the pump in the Mott phase. Similar to Floquet engineered virtual states and bandwidths, such a transient enhancement does not persist at longer times when the interaction balance already has been perturbed by the external field. It is known that both the quasiparticle weight and the interaction strength can affect superconductivity. Thus to understand the differences in dynamics from the Mott and Peierls sides, we evaluate the density of states (DOS) during the pump-probe process [see the Supplementary Material \cite{SUPMAT} for calculation details]. As shown in Fig.~\ref{fig:4}(a), the DOS is gapped in the Peierls phase before the pump, and becomes progressively filled near the Fermi level after the pump. Due to the existence of different instabilities near the phase boundary, these filled weights form another gap-like structure, which as a many-body effect forbids the recovery of the manipulated spin fluctuations. This explains why the transiently photoinduced $\Lambda_d$ could survive after the pump and constantly give rise to the pairing correlation, in contrast to the $u\ll u_c$ case in Fig.~\ref{fig:2}.

To investigate the impact of quasiparticle weight, Figs.~\ref{fig:4}(b) and (c) extract the DOS(0) at various pump strengths, compared with $\langle \Delta_d^\dagger \Delta_d\rangle$ and $\Lambda_d$. Starting from either insulating phases, the pump field always enhances the weights, which could potentially lead to enhanced pairing. Unlike the Peierls phase, the enhancement of DOS(0) in the Mott phase is simultaneously accompanied by the overall drop in spin excitations. These two effects cancel out and the charge fluctuations as well as phonons soon develop, which suppresses remnant $d$-wave superconductivity.

Therefore, the control of spin and charge excitations plays a dominant role in enhancing $d$-wave superconductivity, which is only possible while pumping from the Peierls phase where spin excitations are initially frozen. In contrast to the tiny enhancement due to purely kinetic or quasiparticle reasons [Fig.~\ref{fig:2}(a)], the fluctuations near a quantum phase transition are necessary to obtain considerable enhancement of superconductivity. Note here we discuss only the incipient pairing instability emergent from the competing phases, without implying whether the superconducting ``order'' is dominant in the thermodynamic limit. Although the equilibrium Mott phase displays larger absolute pairing correlation compared to the pumped Peierls phase, the former is known dominant by an SDW order instead of superconductivity.

To summarize, we have examined the nonequilibrium dynamics of a Mott-Peierls system under a pulsed pump and found that the $d$-wave pairing correlations can be enhanced considerably when the original system lies in the vicinity of the phase boundary between CDW and SDW orders. By comparing the dynamical change of pairing susceptibilities with different interactions and fluences, we attribute this enhancement to the manipulation of competing phases and effective interactions near a critical point. More specifically, the enhanced spin fluctuations projected in a $d$-symmetry are consistent with the underlying pairing mechanism. The increase of spin fluctuations provides a pairing glue for superconductivity, which together with photoinduced quasiparticle weight, drives incipient $d$-wave superconductivity. The result indicates that the observed nonequilibrium enhancement of orders or instabilities may originate more from the effective interactions than kinetic or quasiparticle reasons. This study thereby provides an approach to design a photoenhanced state near the critical region where various orders become intimately balanced.

This material is based upon work supported by the U.S. Department of Energy, Office of Science, Office of Basic Energy Sciences, Division of Materials Sciences and Engineering, under Contract No. DE-AC02-76SF00515. Y.W. was supported by the Stanford Graduate Fellowship in Science and Engineering, and the Postdoctoral Fellowship in Quantum Science of the Harvard-MPQ Center for Quantum Optics. C.-C. C. is supported in part by the National Science Foundation under Grant No. OIA-1738698. This research used resources of the National Energy Research Scientific Computing Center (NERSC), a U.S. Department of Energy Office of Science User Facility operated under Contract No. DE-AC02-05CH11231.

\bibliography{paper}
\end{document}